\documentclass[11pt,letterpaper,parskip,bibtotoc]{article}

\usepackage[margin=1in]{geometry} 

\usepackage{setspace}
\usepackage{booktabs}
\usepackage{tabularx}
\usepackage{graphicx}
\usepackage{makecell}
\usepackage{csquotes}
\usepackage{float}
\usepackage{lineno}
\usepackage[super,comma]{natbib}

\title{A new sociology of humans and machines}

\author{Milena Tsvetkova\textsuperscript{a*}, 
Taha Yasseri\textsuperscript{b,c,d}, 
Niccolo Pescetelli\textsuperscript{e,f}, 
and Tobias Werner\textsuperscript{g}}

\date{}

\begin{document}

\maketitle

\vspace{-0.3cm}
\begin{spacing}{1}
\noindent \textsuperscript{a} Department of Methodology, London School of Economics and Political Science, London, United Kingdom

\noindent \textsuperscript{b} School of Sociology, University College Dublin, Dublin, Ireland

\noindent \textsuperscript{c} Geary Institute for Public Policy, University College Dublin, Dublin, Ireland

\noindent \textsuperscript{d} School of Social Sciences and Law, Trinity College Dublin, Dublin, Ireland

\noindent \textsuperscript{e} Collective Intelligence Lab, New Jersey Institute of Technology, Newark, New Jersey, USA

\noindent \textsuperscript{f} The London Interdisciplinary School, London, United Kingdom

\noindent \textsuperscript{g} Center for Humans and Machines, Max Planck Institute for Human Development, Berlin, Germany
\end{spacing}
\vspace{0.3cm}
\noindent \textsuperscript{*} e-mail: m.tsvetkova@lse.ac.uk

\vspace{0.3cm}


\section*{Abstract}

From fake social media accounts and generative-AI chatbots
to trading algorithms and
self-driving vehicles, robots, bots, and algorithms are
proliferating and permeating our communication channels, social
interactions, economic transactions, and transportation arteries.
Networks of multiple interdependent and interacting humans and
intelligent machines constitute complex social systems where the
collective outcomes cannot be deduced from either human or
machine behavior alone. Under this paradigm, we review recent
research and
identify general dynamics and patterns in situations of competition,
coordination, cooperation, contagion, and collective decision-making,
with context-rich examples from high-frequency trading markets, 
a social media platform, an open-collaboration community, and a discussion forum. To ensure more robust and resilient 
human-machine communities, we require a new sociology of humans and machines. Researchers should study these communities using complex-system methods, engineers should explicitly design AI for human-machine and machine-machine interactions, and regulators should govern the ecological diversity and social co-development of humans and machines.

\vspace{1cm}


\noindent Robotic trains and cars drive us around,
auction bots outbid
us for purchases, ChatGPT answers our questions, while social media bots
feed us with dubious facts and news. Modern society is a complex
human-machine social system in which machines are becoming more numerous,
human interactions with machines -- more frequent, and
machine-machine interactions -- more consequential.
With recent advances in generative AI models, the existential threat of 
unexplainable and uncontrollable general AI is looming large again \cite{metz2022,milmo2023,lipton2023}. 
However, when they are numerous and interdependent, even simple 
unintelligent artificial agents can
produce unintended and potentially undesirable outcomes. 
If we want to prevent financial crashes, improve road
safety, preserve market competition, increase auction market efficiency,
and reduce misinformation, it is no longer sufficient to understand
humans -- we need to consider machines, understand how humans and
machines interact, and how the collective behavior of systems of humans
and machines can be predicted. We require a new sociology of humans and machines.

\begin{figure}[h]
\centering
\includegraphics[scale=1]{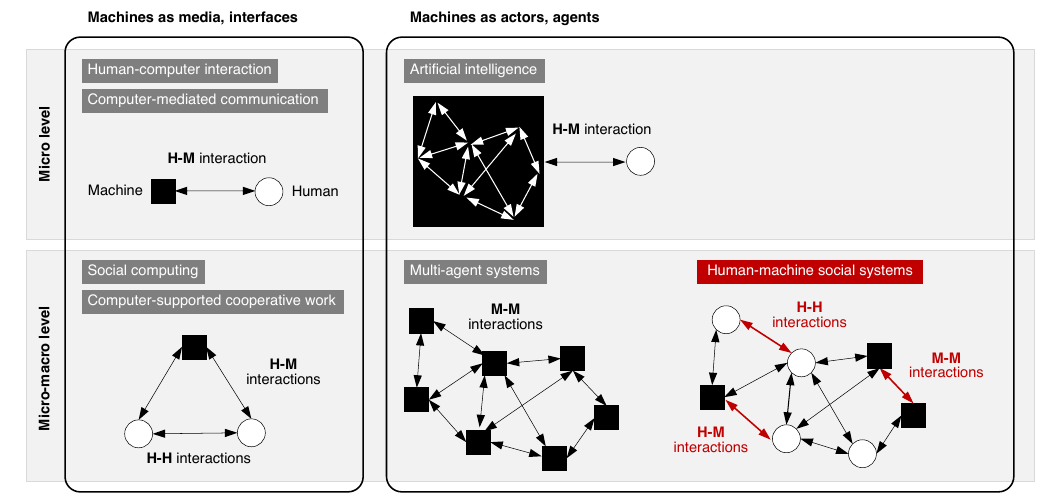}
\caption{Human-machine social systems include
multiple algorithms, bots, or robots that interact among themselves
and with humans in groups and networks. 
Existing fields tend to either approach machines as media or interfaces, 
not autonomous actors or agents, or focus on their cognition and decision-making, 
not group interactions with humans. We call for a new sociology of humans and machines
to study the human behavior, machine behavior,
and the human-human, human-machine, and machine-machine interactions simultaneously in these
complex systems.}
\label{fig:fig1}
\end{figure}

This Perspective synthesizes research and ideas related to social systems
composed of multiple autonomous yet interacting and interdependent
humans and machines such as algorithms, bots, and robots (Fig.\ \ref{fig:fig1}). 
Similar to the conceptualizations of socio-technical systems
\cite{emery2016}, actor-network theory \cite{latour2007,law1992},
cyber-physical social systems \cite{sheth2013,wang2010},
social machines \cite{hendler2010semantic,buregio2013,shadbolt2013}, 
and human-machine networks \cite{eide2016,tsvetkova2017a},
our principal assumption is that humans and machines form
a single social system. In contrast, we do not approach machines as a single
medium or entity -- ``technology'' -- but emphasize their multiplicity, independence, and
heterogeneity and their interactions \cite{cavallaro2009mining}. 
Further, we approach the aggregates as 
complex social systems where network effects and nonlinear dynamic processes
produce collective outcomes that cannot be necessarily deduced from individual preferences and behavior alone\cite{bianconi2023complex}. Our conceptualization extends and complements
the ecological approach to studying machine behavior \cite{rahwan2019},
the ``hybrid collective intelligence perspective'' \cite{peeters2021},
and the budding field of Social AI \cite{pedreschi2023}. We aim to offer
a conceptual overview of the topic. Hence, we do not utilize a systematic 
literature search strategy and instead present selected examples to showcase the new conceptualization \cite{pare2017,sylvester2013,whittemore2005}.

\section*{Human-machine interactions}

We use the term ``machines'' to refer to a range of computational
artifacts: embodied in physical devices such as humanoid robots or existing only in digital space, such as bots and algorithms, and varying in sophistication from simple expert systems that use pre-defined if-else rules to generative deep-learning models that learn from data in real-time. 
Like humans, the machines we consider exhibit
diverse goal-oriented behavior shaped by information and subject to constraints. 
However, the actual cognition and behavior of machines differ from those of humans.
Machines' behavior tends to be predictable and persistent \cite{tsvetkova2017b},
with higher precision and faster execution \cite{hilbert2020a}, 
better informed with access to global information \cite{koren2022}, 
and less adaptable and susceptible to influence
\cite{ferrara2016,ross2019,takko2021}. 
In contrast, humans tend to be limited to local information, satisfice, act with
errors, learn and adapt, succumb to social influence and peer pressure,
yet also exhibit opinion stubbornness and behavioral inertia; on
occasions, they may also use metacognition and revise their own
perceptual and decision-making models. Humans often exhibit cognitive
biases due to limited information processing capacity, bounded
rationality, reliance on heuristics, vestiges of evolutionary
adaptation, and emotional motivations \cite{gilovich2008,kahneman2012},
and algorithms trained on data generated by humans may reproduce
these biases \cite{kordzadeh2022,oneil2016}. Research on
human-like general AI aims to erase the cognitive and
behavioral differences between humans and machines, while work
on human-competitive AI strives for superintelligence that is 
faster, smarter, and more precise than humans' \cite{russell2022,tegmark2017}. 
Either way, humans will remain distinct from machines in the near
future.

Research from the CASA (computers as
social actors) paradigm in psychology emphasizes
that humans treat and respond to machines similarly to
other humans: people
reciprocate kind acts by computers \cite{fogg1997}, treat them as
politely as they treat humans \cite{nass1994}, consider them as
competent, but
also apply gender and racial stereotypes to them \cite{nass2000,siegel2009}; 
people also humanize and empathize with machines, experiencing distress 
when witnessing the mistreatment of a robot \cite{rosenthal-vonderputten2013,slater2006}. 

Nevertheless, there are visible neurophysiological differences in the
brain when humans interact with robots \cite{krach2008,mccabe2001}, 
likely because humans do not
attribute agency and morals to them \cite{gray2007,zhang2022}.
AI is perceived to have lower intentional capacity, lack self-interest, and be
more unbiased than humans.
Consequently, humans exhibit a narrower
emotional spectrum with machines than with humans, reacting 
with lower and flatter levels of
social emotions such as gratitude, anger, pride, and a sense of fairness 
\cite{adam2018,chugunova2022,hidalgo2021,schniter2020},
yet judging machines more harshly when they commit mistakes, 
cause harm, or incur losses \cite{dietvorst2015,candrian2022}.
Further, humans behave more
rationally and selfishly with machines, cooperating and sharing less and demanding
and exploiting more \cite{erlei2022,ishowo-oloko2019,karpus2021,march2021}.
People would design a machine to be more cooperative than they are themselves 
\cite{demelo2019} but act pro-socially towards it only if it is 
more human-like \cite{oliveira2021},
or if it benefits another human \cite{hayes2014}.
Compared to a single person, small
groups of people are even more likely to exhibit competitive behavior
and bullying toward robots \cite{sebo2020}.
Despite this intergroup bias, humans are still susceptible
to machine influence when making decisions or solving problems \cite{kobis2021}. 
Robots can cause both informational and normative conformity
in people \cite{salomons2018,salomons2021} and AI and ChatGPT 
can corrupt humans' moral judgment and make them follow
unethical advice \cite{leib2021,krugel2023}.
Humans tend to trust algorithmic advice more than advice coming from a
another human or a human crowd \cite{bogert2021,logg2019} but may also avoid it
if they perceive a threat to their decision control or a lack
of understanding and cognitive compatibility \cite{burton2020,mahmud2022}.

\section*{Collective outcomes}

\begin{figure}[h]
\centering
\includegraphics[scale=1]{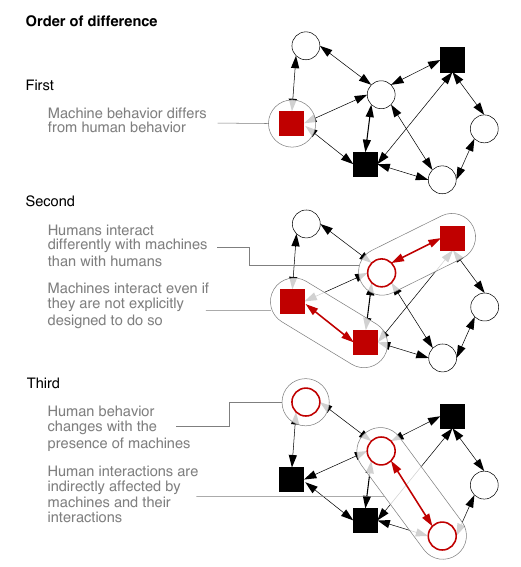}
\caption{Collective outcomes in human-machine social systems differ from
those in human-only systems. Machines behave differently from
humans and in social systems with covert artificial agents, even if humans, 
unaware of the presence of machines, do not change their behavior, the collective outcomes will differ simply because machines act differently. Further, the two types of actors and their interactions are interdependent and influence each other. Thus,  
suspicion or awareness of machine presence 
can change human behavior and interacting with a machine and observing 
machine-machine interactions can influence how humans act toward each other.}
\label{fig:fig2}
\end{figure}

In complex social systems, individuals' behavior and interactions 
affect the collective outcomes, but the relationship can be fundamentally different 
from a simple sum or average \cite{axelrod1984,schelling1971,granovetter1978,miller2009}.
The collective outcomes in human-machine social systems differ from those in 
human-only systems because machines behave differently from humans, human-machine (H-M)
and machine-machine (M-M) interactions differ from human-human (H-H) interactions, 
but also the humans, the machines, and their interactions influence each other indirectly 
(Fig.\ \ref{fig:fig2}). We synthesize common dynamics and patterns 
in groups and networks of humans and machines for five different
social interaction situations (Table \ref{table1}).

\subsection*{Competition}

Competition occurs when multiple actors strive for a common goal that
cannot be shared, as is the case in contests, auctions, and product markets.
Market algorithms are typically designed to benefit the owner without regard for
others or the efficiency and stability of the
market, and yet, they may still benefit the collective. 

With more advanced data processing, learning, and optimization
capabilities than humans, algorithms are better able to discover arbitrage
opportunities and, hence, eliminate mispricing and increase liquidity in
markets. Experimental studies show that algorithmic traders can increase market efficiency \cite{grossklags2006}, but possibly at the expense 
of human traders' performance \cite{angerer2023,cartlidge2012}.
Furthermore, algorithmic traders affect human behavior indirectly:
with their presence, they make human traders act more rationally, and thus, 
reduce strategic uncertainty and confusion in the market \cite{akiyama2017},
reducing price bubbles and bringing prices closer to the fundamental value \cite{farjam2018}.

While perfectly optimizing arbitrage algorithms eliminate mispricings, neither zero-intelligence algorithms that submit random bids without profit maximization \cite{gode1993} nor profit-maximizing agents that update their beliefs from trading history \cite{gjerstad2007} can improve market quality. Meanwhile, manipulator and spoofing algorithms that act to mislead and influence other traders worsen market efficiency \cite{bao2022}. 
Algorithms can also reduce the rationality of professional traders
and alienate and drive away amateur ones.
In an online cryptocurrency
marketplace, traders 
herd after a bot buys, producing larger buying volumes \cite{krafft2017},
while in a Chinese peer-to-peer lending platform, automated investment 
piques inefficient investor scrutiny, increasing bidding duration 
without improving investment return \cite{chen2021}. 
In online auction
markets, naive first-time bidders respond negatively to being outbid by sniping algorithms
and become less likely to return to another auction \cite{backus2015}.
Snipers, which place last-moment bids \cite{roth2002},
work mainly because they exploit the naivety of amateur online
bidders, who tend to increase their bids incrementally. However, 
human lack of rationality has its benefits
because squatting (placing a high early bid) deters new
entrants \cite{ely2009}. In fact, sniping algorithms 
yield a negligibly small \cite{ely2009} or non-existent \cite{gray2013} 
buyer gain, giving them a net negative impact on the marketplace.

In addition to trading and auction markets, pricing algorithms have become widespread in regular product markets \cite{aparicio2023artificial,chen2016empirical} because they either provide recommendations to human pricing managers
\cite{garcia2023strategic, hunold2023algorithmic} or entirely dictate pricing for some firms \cite{chen2016empirical, assad2024}. While pricing algorithms can help firms scale and respond to changes in demand, they may also generate anti-competition. Q-learning algorithms learn to set anti-competitive prices without communication in simulations \cite{calvano2020a,calvano2020b,klein2021, johnson2023platform}, and in experiments, those algorithms are often more collusive than humans in small markets \cite{werner2022}
and foster collusion when interacting with humans compared to fully human markets \cite{normann2023human}. Observational studies of gasoline markets \cite{assad2024} and e-commerce \cite{musolff2022algorithmic, wieting2021algorithms} support the experimental evidence. Furthermore, algorithms can weaken competition by providing better demand predictions, thereby stabilizing cartels \cite{miklos-thal2019, oconnor2021, martin2024demand} or by asymmetries in pricing technologies and commitment \cite{brown2021, leisten2022algorithmic}.

The general intuition is that markets with more actors should be more efficient. Thus, one expects enhanced performance from markets populated by algorithms. However, in reality, the beneficial effects of machines are often in balance, crucially depending on the machines' prevalence, decision speed, and information quality \cite{menkveld2016}, as well as the humans' experience and expectations.

\subsection*{Coordination}

The problem of coordination requires adopting a strategy identical to
or, in some cases, dissimilar from other people's strategies, as when
deciding whether to join a protest, agreeing on a convention such as driving 
on the right-hand side of the road, adopting a communication technology, 
or avoiding a crowd or traffic congestion \cite{ullmann-margalit2015,young1993}. 

In human-machine systems, bots could be used to
introduce more randomness and movement to steer human groups toward
better solutions. Thus, bots acting with small
levels of random noise and placed in central locations in a scale-free network
decrease the time to coordination, especially when the 
solutions are hard to find \cite{shirado2017}. The bots reduce
unresolvable conflicts not only in their direct interactions but also in
indirect H-H interactions, even when the participants
are aware that they are interacting with machines. 
Bots that are trained on human behavior, however, process 
information less efficiently and adapt slower, causing
hybrid groups playing a cooperative 
group-formation game to perform worse than
human-only and bot-only groups  \cite{takko2021}. 

In sum, in situations where a group may get stuck on a suboptimal
equilibrium, non-human-like bots may be able to help by jittering the
system with randomness and unpredictability. Such simple bots may be
more beneficial than bots that superficially imitate human behavior
without the ability to learn and adapt.

\subsection*{Cooperation}

The problem of cooperation pertains to social dilemma situations where a
decision is collectively beneficial but individually costly and risky.
Although the economically rational decision in non-repeated anonymous interactions is to
free-ride and exploit others' contributions, people's actual behavior
tends to be informed by norms of reciprocity, fairness, and honesty
signaling. Thus, as a result of millennia of evolutionary adaptation,
people generally cooperate with each other. If people know they are
interacting with bots, however, they cooperate less \cite{ishowo-oloko2019,karpus2021}. 
Yet, since humans
reciprocate to and imitate cooperative neighbors, introducing 
covert, persistently cooperating bots could increase cooperation.

Computer simulations show that 
persistent prosocial bots favor the emergence of fairness 
and cooperation \cite{santos2019,sharma2023}, 
with stronger effects when humans are more prone to imitation and bots occupy more central
positions in networks with highly heterogeneous connectivity \cite{shen2022}.

Just a few under-cover cooperative bots
can increase cooperation, especially if the bots are widespread in the
network, interacting with many human players rather than
concentrated with overlapping sets of partners \cite{suri2011}.
The reason is that humans wait for someone else to cooperate before they
do, but once they observe many cooperators, they become more likely to
exploit.

Yet, cooperative bots may sometimes fail to improve
cooperation. For instance, hybrid groups with identifiable bots do not
perform better than human-only groups \cite{fernandezdomingos2022}. 
When participants are aware of the presence of artificial agents but not their identity, 
there is a small increase in cooperation
of the bots' direct neighbors but no significant boost in the overall
network \cite{kirchkamp2007}. Similarly, multiple
well-dispersed covert bots, whether all-cooperating or reciprocating, fail to
improve cooperation \cite{shirado2020}, although a single overt network
engineer bot who suggests connecting cooperators and excluding defectors
can successfully do so.

In sum, covert, persistently cooperating
bots (i.e., not very human-like) can increase cooperation in the group 
depending on the network of interactions. Bots are
successful if they are strategically positioned -- well dispersed
in regular and random networks or centrally located in networks with
skewed degree distributions -- or have the power to
strategically engineer the network by offering opportunities to break
links to defectors.

\subsection*{Contagion}

Contagion concerns the spread of
information and behaviors, such as 
memes, slang, fashion, emotions, and opinions
in communication networks \cite{centola2018,christakis2013,rogers2003}. 
In contrast to the
strategic interdependence under the competition, coordination, and
cooperation scenarios, the main mechanism here is
social influence: the tendency to rely on information from others 
to handle uncertainty and
to conform to the expectations of others to fit in society \cite{cialdini2008,deutsch1955,turner1991}.
In human-machine systems, bots can be
remarkably influential at the collective level despite exerting limited
direct influence on individuals because, in networks, small effects can produce chain
reactions and trigger cascades
\cite{fowler2010,leskovec2007,watts2002,pescetelli_bots_2022}.

This is how social media bots influence public opinion. 
In agent-based models of belief formation, weak bots do not alienate their
followers and their followers' friends and thus have their message
spread farther than messages by more pushy and assertive users \cite{keijzer2021}. In
other words, network amplification occurs through bots' indirect
influence precisely because their direct influence on humans is weak,
slow, and unobtrusive. If social media bots influence not people's
opinion but their confidence to express it, they can amplify marginal
voices by triggering the spiral of silence amongst disagreeing humans \cite{ross2019}.
The bots are more influential when they are more
numerous and connected to central actors. 
Strategically placed zealot bots can in fact
bias voting outcomes and win elections \cite{stewart2019}.

Bots can also trigger emotional contagion in
groups, even though they evoke flatter emotional reactions
from individual humans. Humanoid robots can encourage and increase
social interactions among older adults within care facilities, between
different generations, and for children with ASD \cite{sebo2020}. 
In small-team collaborative experiments, a robot's verbal expressions of
vulnerability can show ``ripple effects'' and make the humans more
likely to admit mistakes,
console team members, and laugh together \cite{strohkorbsebo2018},
engage in social conversations and appreciate the group
\cite{traeger2020}. The reported positive contagion effects,
however, were detected when comparing 
one machine to another \cite{traeger2020,zhang2023}.
Overall, bots are more effective than no bots to influence opinions,
behavior, and emotions, but not necessarily more effective than humans.
Yet, even when bots have a weak direct influence on humans' opinions,
they can exert significant collective influence via persistence,
strategic placement, and sheer numbers.

\subsection*{Collective decision-making}

Collective decision-making involves groups making choices or solving problems by combining individual opinions. 
It impacts social phenomena
as diverse as team collaboration, voting, scientific innovation,
and cultural evolution \cite{bang2017}. Originating with Galton's
work on estimation tasks
\cite{galton1907}, the ``wisdom of crowds'' concept suggests that a crowd's aggregated estimate is often more accurate than any individual's, or sometimes even
experts'\ \cite{surowiecki2005}. Crowds perform better
when individual opinions are either independent or diverse
\cite{page2008}, while social interaction can hinder
 \cite{frey2021,lorenz2011,muchnik2013} or improve \cite{becker2017,navajas2018}
collective performance. In human-machine systems,
algorithms introduce diversity and can thus improve decision-making.

An analysis of professional Go players' moves
over 71 years suggests that AlphaGo, the AI program Google
DeepMind introduced in 2016, led human players to
novel strategies and improved their decision-making
\cite{choi2023,shin2021,shin2023}. AlphaGo's
decisions, untethered by human bias, sparked human innovation in this game. However, the positive influence of machine-human social learning on problem-solving may be limited. When algorithms are introduced in chains of humans
engaged in sequential problem solving, the innovative solutions benefit immediate followers, but
team accuracy
does not have lasting effects because humans are more likely to replicate human solutions than algorithmic ones \cite{brinkmann2022}. Similarly, in a team prediction task, an algorithm maintaining group diversity by promoting minority opinions improves individual accuracy, but the effects dissipate for team accuracy \cite{pescetelli_variational_2022}.

The area of hybrid intelligence investigates how and when to combine human and algorithmic
decision-making \cite{dellermann2019,peeters2021} and includes
research on active and interactive learning and human-in-the-loop
algorithms \cite{wiethof2021}, with applications in
clinical decision-making, where
combining clinician and algorithmic judgments can improve
cancer diagnoses \cite{hekler2019,tschandl2020}, and citizen science, where combining crowd-based with machine classifications can improve accuracy. On Zooniverse, a prominent citizen science platform, this hybrid approach found supernovae candidates among Pan-STARRS images more effectively than humans or machines alone \cite{wright2017} but damaged
citizen scientists' retention \cite{bowyer2015,trouille2019}, suggesting a trade-off between efficiency and volunteer engagement. Ultimately, the deployment of machines could further marginalize certain groups of volunteers \cite{ibrahim2021gender}, and with fewer volunteers, AI's performance could diminish.

The emerging field of hybrid intelligence suggests that algorithms introduce novel solutions, but these may be too unfamiliar for humans to adopt. Nevertheless, machine diversity and competition might inspire alternative forms of human creativity and innovation. Developing methods to effectively combine human and machine solutions could further improve collective intelligence \cite{cui2024ai}.

\section*{Implications}

The algorithms in current human-machine social systems are
relatively simple. Few use sophisticated machine learning
or AI, and typically, these guide narrow and specific
behaviors \cite{yang2023,yang2024}. 
Except for malicious social
bots and customer-service chatbots, most machines do not mimic
human qualities. Most machines are superhuman ---
processing vast data amounts, acting swiftly, and
handling tedious tasks--- or candidly non-human--- resisting peer influence, not reciprocating, and acting randomly. There is a clear distinction between covert
and overt bots: covert bots are more problematic 
than bots declaring
their identity and following norms and regulations. 

The effects of machines on human-machine social
systems vary by their number, algorithms,
network position, interaction situation, institutional
regulations, technological affordances, organizational context, and
emerging norms and culture (see Boxes 1-4). Machines alter outcomes through their
unique behavior because humans interact differently with them, and because of their indirect effects -- machines' presence changes how
humans interact amongst themselves.

Machines can be beneficial when they act or steer humans to counteract human weaknesses. For instance,
noisy bots can disrupt sub-optimal outcomes and improve coordination,
persistently cooperative bots can curb retaliation and
maintain cooperation, machines in central roles as
arbitrageurs improve price discovery and market quality, and network-engineering bots boost collective welfare via cooperator assortment and defector exclusion.
With global information, higher processing power, and
instantaneous execution, machines can quickly address external events
like vandalism or political and natural crises, ensuring
system robustness, resilience, and efficiency.
Depending on the situation, machines offer superhuman persistence or randomness, 
norm-setting rationality, or
solution diversity, enhancing human behavior
towards better outcomes.

However, what helps in one context can hinder in another. Machines'
unintuitive solutions may confuse humans, hindering innovation and technological progress.
Humans might not act fast enough to correct
machines' errors, resulting in instabilities and flash
failures. Machines are less adaptive than humans to changes,
impeding system evolution. Machines can be designed
to exploit human weaknesses, triggering
cascades that exacerbate polarization, emotional contagion,
ideological segregation, and conflict. Machines' non-human optimization
logic, execution speed, and behavioral rigidity can clash with human behavior, pushing interactions toward undesirable outcomes.

\subsection*{Research}

Existing research is often biased towards engineering and optimization,
lacking deeper insights from a social science perspective. The time for a
new sociology of humans and machines is critical, 
before AI becomes more sophisticated: generative AI exhibits
emergent behavior that itself requires explanation
 \cite{ray2023,webb2023}, complicating the understanding of system
dynamics. 

Researchers would benefit from an agent-based modeling framework that outlines
distinctions between human and bot agents: utility function, optimization
ability, access to information, learning, innovation/creativity,
accuracy, etc. The framework could borrow concepts from other two-agent systems, such as
predator--prey, principal--agent, and common pool resource models. 
Controlled experiments should explicitly compare human-machine, human-only and machine-only
networks, and known bots against covert bots. Experiments could
manipulate participants' perceptions of algorithms'
technical specifications, agenthood \cite{frey2014}, emotional capability,
and biases. Field interventions in online communities with endemic bot populations present another promising direction. Existing examples include social bots that gain influence by engaging human users
\cite{aiello2012,freitas2015,messias2013,savage2016}, trading bots that
manipulate prices in cryptocurrency markets \cite{krafft2018}, 
political bots that promote opposing political views to decrease polarization \cite{bail2018},
and ``drifters'' to measure platform bias \cite{chen2021}.
Expanding on the cases reported here, we need observational research on additional
human-machine communities and contexts such as traffic systems
with human-driven and driverless vehicles, 
online multiplayer games comprising human players, non-player
characters, and cheating code, and dating markets with AI-driven chatbots
 \cite{lorenz2023}.

Finally, research with artificial agents introduces ethical problems demanding careful elaboration and mitigation. 
Research protocols should minimize 
interventions \cite{salganik2019}, possibly deploying covert bots only 
where they already exist, ensuring their actions are not
unusual or harmful \cite{krafft2017}. Even then,
bots may still face opposition from users due to privacy concerns  \cite{aiello2012}. Do people perceive certain
bots as inherently deceptive? Could knowledge of the bot owner and
algorithm mitigate this perception?

\subsection*{Design}

Humans are resilient and successful when they exhibit high levels of efficient communication, context awareness, emotional recognition and response, and ethical and cultural sensitivities. These features should be encouraged when designing social machines to build trust, ensure legal compliance, promote social harmony, 
enhance user satisfaction, and achieve long-term sustainability \cite{paiva2018towards,wagman2021beyond}. 

Since H-H, H-M, and M-M interactions
differ, machines should be specifically designed for
each scenario, with separate training for each interaction. This could avoid market underperformance 
when bargaining algorithms trained on human-only markets 
adapt poorly to human-machine negotiations \cite{erlei2022},
or traffic jams when driverless vehicles trained on human
driving fail to properly interact with each other \cite{chang2023}.

AI design should also adopt a hierarchy of behavioral rules and
conventions guiding H-M and M-M interactions in the context of H-H
interactions. Isaac Asimov's famous Three Laws of Robotics  \cite{asimov2004}
which regulate M-H interactions and self-preservation---1)
not harming humans, 2) obeying humans, and 3) protecting own existence, with priority given to
higher order rules---could be adapted for M-M interactions, considering specific contexts, implications, and unintended consequences \cite{graham2017}.

Further, cultural context is crucial for AI design. 
People's perceptions of machines vary by age, environment, personality, and geography  \cite{awad2018,hidalgo2021}. 
Machines reflect their developers' culture and operate in settings with specific 
organizational and community norms \cite{tsvetkova2017b}.  
Thus, AI design for self-driving cars and assistant bots should consider local driving culture 
and attitudes toward domestic assistants during training.

Finally, the systems we reviewed exhibit complexity in that their emergent behavior transcends a mere aggregation of individual components, displaying what is known as ``network effects.'' While natural complex systems often demonstrate remarkable resilience and adaptivity \cite{pinheiro2017university}, the human-designed complex systems discussed here, albeit to varying degrees, are not inherently adaptive. To enhance resilience and robustness, AI designers should incorporate complex adaptive system principles, like negative feedback, modularity, and hierarchical organization. For example, dense networks lacking diversity and modularity are susceptible to systemic failures \cite{may2008,balsa-barreiro2020,bak-coleman2021} and are easy to control \cite{liu2021}. Introducing bots that increase network diversity, introduce resistance, build resilience, or incorporate negative feedback could enhance adaptability and stability. Such configurations have improved group outcomes in forecasting \cite{pescetelli2021}, exploration/exploitation \cite{mason2012}, and general knowledge tasks \cite{navajas2018}. Literature on monitoring and steering complex systems can help predict critical transitions  \cite{scheffer2012,centola-becker2018,bianconi2023complex}, designing safer human-machine systems across different environments and perturbations. 

\subsection*{Policy}

We urge a system-focused approach to AI policy and ethics: policymakers should approach AI not as a single existential threat but as a multiplicity of machines and algorithms.
Machines are often more beneficial when they are superhuman or simply ``alien''
and when they are diverse.
Similarity in information sources,
interaction speeds, optimization algorithms, and objective functions can
cause catastrophic events, like flash crashes in markets. 
Thus, while AI designers may chase optimization and superintelligence, policymakers
should focus on the diversity of human-machine ecologies. 
Policymakers should demand adaptivity and resilience, too. 

Policymakers should also anticipate the social co-development of
machines and humans, which will inadvertently change
existing institutions. Machines can cause humans to
withdraw interaction: for instance, outsourcing care to robots reduces caregivers' empathy  \cite{kenway2023}. Intelligent machines are changing the transmission and creation of human culture, altering social
learning dynamics, and generating new game strategies, scientific
discoveries, and art forms  \cite{brinkmann2023}. Humans must
adapt to intelligent machines just as intelligent machines must
learn from and adapt to humans. Finally, ethicists should address questions such as: Should all machines be equal? 
Should we allow status hierarchies, possibly reflecting and exacerbating 
existing socio-economic inequalities?

\section*{Conclusion}

This Perspective synthesizes relatively
disparate literature 
based on agent-based models, controlled experiments, online field interventions, 
and observational analyses from human-computer interaction,
robotics, web science, financial economics, and computational social
science under a common theoretical framework: human-machine social
systems. We identify common dynamics and patterns that
emerge from the interactions of humans and intelligent machines
regardless of the specific context, as well as peculiarities and unique
problems that concrete techno-organizational and socio-cultural
environments generate. Our utmost ambition is to stimulate cumulative
empirically driven and mechanism-focused sociological research in the
emerging, fast-evolving field of human-AI science. At stake are new and
urgent social challenges such as online misinformation, market flash
crashes, cybersecurity, labor market resilience, and road safety. With
increasing social connectivity and accelerating developments in AI,
understanding the complex interactions between humans and intelligent machines is a
challenging undertaking, but one that is crucially important for a
better human future.

\section*{Acknowledgments}

MT acknowledges the generous support of the 
Santa Fe Institute during the period when the research was conducted. TY was partially funded by the Irish Research Council under grant number IRCLA/2022/3217,
ANNETTE (Artificial Intelligence Enhanced Collective Intelligence).
The funders had no role in study design, data collection and analysis, 
decision to publish, or preparation of the manuscript. 

\bibliographystyle{naturemag}
\bibliography{hmss}

\section*{Author contributions}

M.T. conceived the study and prepared the initial draft. All authors wrote and revised the manuscript.

\section*{Competing interests}

The authors declare no competing interests.

\clearpage

\clearpage

\clearpage

\begin{table} \footnotesize \centering 
    \centering
    \caption{Types, examples, and collective outcomes of human-machine social systems. Boxes 1-4 present more context for four of the examples: high-frequency trading markets, Twitter, Wikipedia, and Reddit. These four communities are clearly defined, relatively large, and well-studied and exemplify situations of market competition, contagion in political communication, cooperation and coordination, and collective action, respectively.} 
    \label{table1} 
    \begin{tabularx}{\textwidth}%
        {>{\raggedright\arraybackslash}p{1.7cm}%
        >{\raggedright\arraybackslash}p{2.5cm}%
        >{\raggedright\arraybackslash}p{2.3cm}%
        >{\raggedright\arraybackslash}p{3.5cm}%
        >{\raggedright\arraybackslash}p{4.5cm}%
        }
        \toprule
        \textbf{Situation} &  \textbf{Models} &  \textbf{General examples} &  \textbf{H-M examples} & \textbf{H-M collective outcomes} \\ 
        \midrule
        Competition & \parbox{2.5cm}{\raggedright Zero-sum game\\Auction markets\\ Buyer-seller markets\\ Oligopoly market games} & \parbox{2.3cm}{\raggedright Competitions\\Contests\\Auctions\\Markets} & \parbox{3.5cm}{\raggedright High-frequency trading markets\\Pricing algorithms\\Online auctions with sniping algorithms\\Cheating bots in multiplayer games} & \parbox{4.5cm}{\raggedright + Increase efficiency by improving liquidity and price discovery\\ -- Increase volatility by causing price spikes and crashes\\ -- Increase in consumer prices from algorithmic collusion\\ -- Decrease human activity} \\
        \midrule
        Coordination  & \parbox{2.5cm}{\raggedright Coordination game\\Stag Hunt game\\Battle of the Sexes\\Chicken game\\Graph Coloring game} & \parbox{2.3cm}{\raggedright Conventions\\Technological standards\\Communication technology\\Transportation\\Supply chain management} & \parbox{3.5cm}{\raggedright Traffic with autonomous vehicles\\Wikipedia editors\\Humanoid robots in warehouses and manufacturing} & \parbox{4.5cm}{\raggedright + Improve coordination by introducing random behavior\\ -- Worsen coordination by failing to adapt} \\
        \midrule
        Cooperation & \parbox{2.5cm}{\raggedright Prisoner's Dilemma\\Public Goods game\\Dictator game\\Ultimatum game\\Trust game} & \parbox{2.3cm}{\raggedright Team collaboration\\Mutual aid} & \parbox{3.5cm}{\raggedright Caring robots\\Personal AI assistants\\Chatbots\\Wikipedia editors\\Reddit moderators\\Non-player characters in multiplayer games} & \parbox{4.5cm}{\raggedright + Increase cooperation\\+ Increase efficiency by handling large task volumes\\+ Increased forecasting accuracy\\-- Decrease efficiency by introducing new types of workload} \\ 
        \midrule
        Contagion & \parbox{2.5cm}{\raggedright Epidemiological models\\Threshold models of contagion} & \parbox{2.3cm}{\raggedright Communicative disease\\Information\\Innovations} & \parbox{3.5cm}{\raggedright Twitter\\Reddit} & \parbox{4.5cm}{\raggedright -- Increase spread of misinformation, opinion polarization, verbal conflict\\+ Increase human activity and engagement} \\
        \midrule 
        Collective decision making & \parbox{2.5cm}{\raggedright Vote aggregation\\Active learning} & \parbox{2.3cm}{\raggedright Crowdsourcing\\Prediction markets\\Voting systems\\Cultural evolution} & \parbox{3.5cm}{\raggedright Clinical diagnosis\\Citizen science\\Content moderation\\Hybrid forecasting} & \parbox{4.5cm}{\raggedright + Increase innovation and accuracy by introducing diversity\\ -- Decrease human activity and engagement} \\
     \bottomrule
    \end{tabularx}
\end{table}

\clearpage

\subsection*{Box 1: Competition in high-frequency trading markets}

High-frequency trading (HFT) algorithms constitute automated scripts that
rely on high-speed, large-volume transactions to exploit mispricings or
market signals before they disappear or are incorporated into the price \cite{hagstromer2013}.
The phenomenon started in the mid-90s and has since spread to dominate
foreign equities, foreign exchange, commodities, futures, and stock
markets globally \cite{menkveld2016}. 

HFT algorithms process large amounts of trade history
data and current news to make decisions and are thus considered the
better ``informed'' traders \cite{brogaard2014}. Some of the
algorithms appear to anticipate the market, and their trades consistently
predict future order flow by human traders \cite{hirschey2021}. However,
since most algorithms react similarly to the same public information,
they exhibit less diverse trading strategies and more correlated actions
among themselves compared to humans \cite{chaboud2014}. Thus,
although their behavior generally improves market efficiency, it can
also trigger behavioral cascades and instability \cite{jarrow2012}.

HFT algorithms generally act as market makers,
increasing trading opportunities, reducing transaction costs, connecting
buyers and sellers across venues, and submitting significant volumes of
price quotes \cite{hendershott2011,menkveld2016}. They facilitate
price efficiency by trading in the direction of permanent price changes
but opposite temporary price errors \cite{brogaard2014}. This
regularly acts as a stabilizing force, reducing 
short-term volatility \cite{chaboud2014,hasbrouck2013,hagstromer2013}.
On a longer time scale, however, HFT algorithms may decease market quality by increasing volatility \cite{boehmer2012} and uncertainty \cite{hilbert2020a}, and by reducing trading strategy diversity \cite{johnson2013}. For instance, although the algorithms did not cause the 2010 flash crash, 
they exacerbated it by amplifying the volatility
\cite{kirilenko2017}. This has led to recent efforts to regulate the speed of trading in markets, for example, by processing trades in batches at slower intervals to diminish the advantage that HFT algorithms have \cite{budish2015high}.

\clearpage

\subsection*{Box 2: Contagion on Twitter}
 
Social bots on the micro-blogging platform Twitter (re-branded as X in 2022) are covert automated accounts 
designed to impersonate humans to boost followers, disseminate information, 
and promote products. Bots and bot detection methods have co-evolved, resulting in increasingly more sophisticated imitation or detection strategies \cite{beskow2018,cresci2020,davis2016,duh2018,ferrara2016,orabi2020,varol2017}, but detection is inherently limited due to the overlap between covert autonomous bots, managed user accounts, hacked accounts, cyborgs, sock-puppets, and coordinated botnets \cite{abokhodair2015,bastos2019,chu2010,grimme2017}. Estimates suggest that 9-15\% of Twitter users are bots \cite{chu2012,varol2017},  with bot activity typically increasing around controversial political events \cite{stella2018}.

Twitter social bots, 
who do not follow social instincts but neither succumb to fatigue, engage less in
social interactions via  replies and mentions 
than humans but produce more content \cite{pozzana2020}. 
The bots mainly retweet -- a passive strategy to indicate support
 and gain followers -- but are less successful in attracting
friends and followers than humans \cite{varol2017}.
Overall, they are less connected and bot-bot (2\%), bot-human (19\%),
and human-bot (3\%) interactions are considerably 
less common than human-human interactions (76\%) \cite{stella2018}.

Despite their rudimentary social behavior and weak network integration, 
Twitter bots significantly influence political communication, 
public opinion, elections, and markets.   
They play an important role in misinformation dissemination in relation to political events \cite{ferrara2017,forelle2015,howard2018,howard2016,shao2018,suarez-serrato2016,yan2023}, 
COVID-19 \cite{himelein-wachowiak2021,yang2020}, and stock market investment \cite{fan2020}. Bots can affect human interaction networks by encouraging followings and conversations \cite{hwang2012} and
amplify low-credibility content early on by targeting influential humans
through replies and mentions \cite{shao2018}. Bots' large numbers
enhance their visibility and influence to trigger deep information cascades \cite{stella2019}. 
Bots equally link to true and false news from low-credibility sites, but people 
prefer false content, making humans ultimately responsible for the spread of false news \cite{vosoughi2018}.

Twitter bots significantly contribute to negative sentiment and conflict escalation. Acting from the periphery, they target central human users to exert indirect influence. They amplify existing negative sentiment and selectively promote inflammatory content, often targeting only one
of the factions \cite{stella2018}. Their success stems from exploiting human tendencies to connect with similar others and engage with messages that reinforce their beliefs \cite{gorodnichenko2021}. Consequently, bots increase ideological polarization and negatively affect democratic discourse on social media, as seen in the 2016 US presidential election \cite{bessi2016}, the 2016 UK Brexit Referendum
\cite{gorodnichenko2021}, and the 2017 Catalan independence referendum \cite{stella2018}.

In sum, Twitter's covert social bots are considered harmful, 
prompting the platform to cull them \cite{bbcnews2018,dang2022}. 
Their strength lies in indirect action: they skew the platform's recommendation system 
to bias content popularity \cite{pescetelli_bots_2022} and 
exploit human behavioral weaknesses like attention seeking, confirmation bias, 
moral outrage, and ideological homophily.

\clearpage

\subsection*{Box 3: Cooperation and coordination on Wikipedia}

Wikipedia, the largest and most popular
free-content online encyclopedia, hosts an ecology of bots
that generate articles, fix errors on pages, link to
other sites and databases, tag articles in categories, identify vandals,
notify users, and so on \cite{halfaker2012,niederer2010,zheng2019}. 
These bots are open-source, documented, approved, registered, and tagged \cite{geiger2011,halfaker2012}. 
They are not sophisticated: most use basic regular expressions or
straightforward heuristics, and only some incorporate machine learning
techniques. 
They are significantly less numerous than human editors but 
complete a disproportionately large volume of all edits
\cite{steiner2014,tsvetkova2017b,zheng2019}. 
Compared to human-human interactions, bot-bot interactions
are more reciprocal and balanced
but do not exhibit status effects
\cite{tsvetkova2017b}.  
While bots are more likely to be involved in back-and-forth reverts with each other
over long periods, these accidental encounters
rarely indicate direct opinion conflict but constitute routine
productive maintenance work or reflect
conflicts existing between their human owners \cite{geiger2017}. Human editors interact mainly
with policing bots, primarily by criticizing 
the legitimacy of the norms they enforce, rather than the sanctions themselves,
suggesting that editors perceive bots as extensions of their
human owners rather than independent agents \cite{clement2015}.

Bots have been invaluable to the maintenance and operation of
Wikipedia. The diversity of the bot ecology guarantees the system's robustness and resilience.
For instance, during the random outage of the anti-vandalism ClueBot NG,
the website eventually caught up, albeit more slowly than usual, thanks to the
heterogeneity of the quality control network, comprising instantaneous
fully automated robots, rapid tool-assisted humans (cyborgs), humans
editing via web browsers, and idiosyncratic batch scripts \cite{geiger2013}. 
Wikipedia demonstrates that successful bot governance and
regulation does not have to sacrifice distributed
development and diversity. The algorithmic simplicity, independence, and heterogeneity of the machines facilitate the system's success and
resilience overall but may also introduce unexpected complexities and
uncertainties in communication at smaller scales \cite{hilbert2020b}.

\clearpage

\subsection*{Box 4: Cooperation and contagion on Reddit}

Reddit is a popular news aggregation, content rating, and discussion website
founded in 2005. Bots on Reddit provide internal
moderation and communication, augment functionality (e.g., for mobile
users), or post content, ranging from comic and
playful posts by evident automated accounts such as haiku\_robot and
ObamaRobot, to trolling and provocative comments by undercover social
bots \cite{massanari2016}. 
Similarly to Wikipedia, Reddit has developed
norms and protocols for deploying bots \cite{hurtado2019}, but similarly to Twitter, it has
no effective service limitations to prevent covert and malicious
automated accounts. Nevertheless, Reddit differs from Twitter in several
crucial ways -- content on the site is posted within communities, heavily moderated,
up-/downvoted, and extensively discussed. Due to these structural
differences, political misinformation, polarization, and conflict are
less pronounced on the platform. 

Reddit offers evidence for collaboration and contagion
between humans and bots. Content moderators have widely adopted 
Automod -- a bot that is flexible to updates,
adaptable to community rules, and interpretable. The bot aids with menial tasks
but does not necessarily decrease workloads as it requires 
continuous updates in response to changes in user behavior and language 
and involves high volumes of correspondence with incorrectly banned users \cite{jhaver2019}. 
On the other side, regular users engage with evident entertainment bots, and 
their direct replies imitate the sentiment and the words of the bot posts \cite{ma2020}. 
Thus, emotional contagion and lexical entrainment can occur between humans and bots,
even when humans are aware of the simple automated script behind the bot.

\end{document}